\documentclass[pra,twocolumn,floatfix]{revtex4}
\usepackage{epsfig}
\begin{document}
\title{An Intense Source of Cold Rb Atoms from a Pure 2D-MOT}
\author{J. Schoser, A. Bat\"ar, R. L\"ow, V. Schweikhard, A. Grabowski, Yu.B. Ovchinnikov, and T. Pfau}
\affiliation{5.Physikalisches Institut, Universit{\"a}t Stuttgart,
Pfaffenwaldring 57, 70550 Stuttgart, Germany}
\date{\today}
\begin{abstract}
We present a 2D-MOT setup for the production of a continuous
collimated beam of cold ${^{87}}$Rb atoms out of a vapor cell. The
underlying physics is purely two-dimensional cooling and trapping,
which allows for a high flux of up to ${6\cdot10^{10}}$ atoms/s
and a small divergence of the resulting beam. We analyze the
velocity distribution of the 2D-MOT. The longitudinal velocity
distribution of the atomic beam shows a broad
feature(FWHM$\simeq$75\,m/s), centered around 50\,m/s. The
dependence of the flux on laser intensity, geometry of the
trapping volume and on pressure in the vapor cell was investigated
in detail. The influence of the geometry of the 2D-MOT on the mean
velocity of the cold beam has been studied. We present a simple
model for the velocity distribution of the flux based on rate
equations describing the general dependencies.
\end{abstract}
\maketitle


Cold atomic beams are needed for many applications in atom optics
\cite{sigel94,metcalf} or in the field of atomic clocks based on
atomic fountains. Especially the process of evaporative cooling
\cite{ketterledruten} demands a high atom number as a starting
point for reaching quantum degeneracy. This requires an intense
source which can efficiently load a magneto-optical trap (MOT) fast.\\
For this purpose sources with a small divergence of the atomic
beam are favorable. The longitudinal velocities in the beam should
be within the capture range of the 3D-MOT. As long as the cross
section of the beam is smaller than the spatial capture range of
the MOT the figure of merit for optimized loading into a
MOT is the total integrated flux up to its capture velocity.\\
Besides background gas loading and pulsed loading mechanisms like
chirped slowers or double-MOT systems many experiments use Zeeman
slowers as a continuous source \cite{zeemanslower}. A
Zeeman-slower decelerates an intense thermal atomic beam along the
propagation axis of the beam by radiation pressure, while the
spontaneous emission processes give rise to a transverse heating
of the atoms. This emerges in a strongly diverging beam with a
flux of up to ${10^{11}}$ atoms/s. In some cases the slowing light
on axis or the magnetic fields involved disturb
the consecutive MOT.\\
Lu et al.\,\cite{LVIS96} have realized a low velocity intense
source (LVIS) of atoms by creation of a dark channel in one of the
six MOT-laser-beams. Due to the imbalance in the radiation
pressure along one axis a continuous beam of cold atoms is coupled
out of a trapped cloud. This source provides a small thermal
background and a narrow velocity profile at velocities below
20\,m/s. However it does not reach the high flux
of a Zeeman slower.\\
A new approach is 2-dimensional magneto-optical cooling and
trapping. This technique has first been used to extract very slow
atoms out of a cooled atomic beam by Riis et al. \cite{riischu}
and was later refined to produce a beam of slow atoms with
velocities from 2 to 10\,m/s \cite{swanson,riis}. An alternative
is to collimate and compress a slow atomic beam like realized by
\cite{nellessen,Yu}. Finally it has been demonstrated that an
atomic funnel, applied in different configurations, produces a
beam of cold atoms out of a vapor cell
\cite{berthoud,dieck98,camposeo}. Dieckmann et al. have realized a
2D$^{+}$-MOT which produces a total flux of ${9\cdot10^{9}}$
Rb${^{87}}$-atoms/s. The plus-sign stands for a pushing laser beam
on axis copropagating with the atomic beam. This setup uses small
cooling laser powers and retroreflection of the laser beams. The
vapor pressure reaches up to saturated vapor
pressure (${\simeq10^{-7}}$\,mbar) at room temperature.\\
Pure 2D cooling does not require light on the axis of the atomic
beam compared to a Zeeman slower, an LVIS or a 2D$^{+}$-MOT. Hence
there is less disturbance of the consecutive 3D-MOT setup. Radial
cooling collimates the outgoing beam while the allowed cooling
time, which depends on the geometry of the cooling region,
influences the possible longitudinal velocities of atoms in the
beam. The resulting flux is comparable to the one of a Zeeman
slower. Moreover it offers a small beam-divergence. In our system
we could achieve a flux of 6$\cdot$10${^{10}}$\,atoms/s with a
beam-divergence of 32\,mrad. The figure of merit characterizing
the performance of our system is the integrated brightness which
is about ${2\cdot10^{12}}$\,atoms/(s$\cdot$srad). On the other
hand one needs to accept a rather high mean velocity of the cold
beam and a relatively broad velocity distribution depending on the
chosen geometry. The mean velocity of the atoms in our apparatus
is $\simeq$50\,m/s. We plan to capture most of the atoms up to a
capture velocity of 60\,m/s by a large volume, elongated 3D-MOT in
a UHV-chamber which the atomic beam enters at a small angle with
respect to the long axes of the elongated MOT.\\

This paper is organized as follows: in section \ref{theory} we
explain the basic principle of a two-dimensional MOT. We discuss
the influence of geometry, laser power and pressure in the vapor
cell on the atomic beam. Based on rate equations we derive a model
for the longitudinal velocity distribution. Section
\ref{experiment} gives a detailed description of the setup and the
measurement techniques used for the characterization. We present
the results and discuss the applicability of our model in part
\ref{results}. We summarize and give an outlook on future steps in
our experiment in section \ref{outlook}.\\

\section{Principle of Operation and Theoretical Model}
\label{theory} The basic geometry of the experimental setup (see
Fig.\,\ref{setup}) is given by a vapor cell separated from a
UHV-chamber by a differential pumping tube which is also the
aperture for the output beam of cold atoms. A two-dimensional
magnetic quadrupole field is produced by four coils with
rectangular shape. The axis of both, the cell and the tube,
coincide with the line of zero magnetic field. Four
perpendicularly counterpropagating laser beams with orthogonal
circular polarisations in the usual MOT-configuration enclose a
cooling volume along the axis of zero magnetic field (z-axis). The
center of the MOT laser beams is positioned about 4\,cm in front
of the differential pumping tube. The cooling volume extends to
the entrance of the tube. Atoms in the vapor cell which enter the
cooling volume are slowed down in the two radial dimensions and
are compressed to the z-axis. The atoms' velocity in the
longitudinal direction is not cooled. Hence the atoms experience a
skew trajectory into the center of the 2D-MOT, according to the
equation of motion of a damped harmonic oscillator, while
propagating along the axis. This produces a thin, dense, well
collimated atomic beam in two directions, namely in the positive
and negative z-direction. We constrict ourselves to one beam
travelling in the positive z-direction, where it passes through
the differential pumping tube. The other beam is lost after it hit
the back
wall of the glass cell.\\
The atoms in the vapor cell have to fulfill three criteria
simultaneously in order to contribute to the flux of laser-cooled
slow atoms at
the exit of the differential pumping tube:\\
a) The initial radial velocity component needs to be smaller than
the transverse capture velocity of the 2D-MOT.\\
b) The interaction time of the atoms with the light field needs to
be long enough so that the trajectory of the atom hits the
entrance of the differential pumping tube (the radial velocity is
sufficiently cooled) such that its
divergence is small enough to make it to the exit of the tube.\\
c) The mean free path in the vapor cell should be larger or
comparable to the length of the 2D-MOT, so that the cooling of the
atoms has not suffered from collisions.\\
At first we consider the collision-less regime. Here the density
is low enough that the mean free path in the vapor cell is larger
than the dimensions of the cell. In this case we can assume that
the thermal atoms start on the walls and no collisions take place
in the volume. The geometry of the tube and the glass cell is
designed in such a way that the opening angle of the differential
pumping tube does not accept atoms starting on the side walls
without being transversely cooled. In this configuration thermal
atoms are only transmitted if they start on that part of the back
wall of the glass cell which lies within the acceptance angle of
the differential pumping tube. This choice limits the thermal background.\\
For an understanding of the total flux and its dependence on the
geometry, the longitudinal velocity profile of the atomic beam
needs to be investigated. This information is given by the range
of accepted velocity classes and its dependence on external
parameters. The concept of a capture velocity, well-known from
3D-magneto-optical-traps, must be modified to embrace the
2D-MOT-configuration. Since the cooling is restricted to
transverse velocities, we define a radial capture velocity. In
spherical 3D-MOTs the capture velocity has a constant value
depending only on detuning, laser beam size and intensity and on
the magnetic field gradient. However, in a 2D-MOT a finite cooling
time is necessary to collimate the atoms onto the beam axis. Atoms
which travel too fast along the z-direction can not be
sufficiently cooled and are filtered out by the aperture. A main
parameter is therefore the cooling time ${\tau=z/v_z}$ which is
given by the longitudinal velocity ${v_z}$ and the distance $z$
from the tube at which the atom enters the cooling volume of the
2D-MOT. Thus the effective transverse capture velocity becomes a
function of $z$ and ${v_z}$. For small ${v_z}$ the radial cooling
time dominates and the capture velocity should be simply a
constant ${v_{c0}}$ determined by the parameters of the cooling
laser beams and the magnetic field gradient. For larger ${v_z}$
the capture velocity should fall off as ${1/v_z}$. The velocity
range between these two asymptotes depends additionally on the
position ${z}$ where atoms enter the cooling volume. With
increasing cooling length one should expect more atoms with high
axial velocities to be transversely cooled
and transmitted through the tube.\\
The cooling time $\tau$ determines the final transverse velocity:
${v_{r,i} - v_{r,f}
=\frac{1}{m}\int_0^{\tau}\,F_{sp}(x(t),v(t))dt}$. Here ${F_{sp}}$
is the spontaneous scattering force, ${v_{r,i}}$ the initial,
${v_{r,f}}$ the final radial velocity and $m$ the mass of a
Rb-atom. By that the divergence of the atomic beam (if it is not
limited by the acceptance angle of the tube) is shaped depending
on the mean length of the cooling volume. Radial cooling within a
finite cooling time increases the portion of atoms with small
v$_z$ in the beam. \emph{This is the reason why the velocity
distribution of the atomic beam is non-thermal and is shifted to
much lower values than a thermal distribution} at room temperature
(${\langle\,v\,\rangle}$$\simeq$275\,m/s) although there is no
longitudinal cooling. A typical time scale for the cooling time is
1-2\,ms. As will be described in part \ref{experiment}, a usual
length of the cooling volume in our setup is about 60\,mm. This
gives a rough estimation for the mean velocity in the atomic beam
between 30 and 60\,m/s.\\
The influence of the MOT-laser intensity is reflected in the
efficiency of the transverse cooling. The capture velocity should
increase with intensity and saturate at some value when the
saturation parameter dominates the spontaneous force.\\
The vertical beam size, in combination with detuning and magnetic
field gradient, determines the radial capture velocity ${v_{c0}}$.
An increase of the length of the 2D-MOT increases the upper limit
of the longitudinal velocities which are trappable. Additionally,
the size controls the mean longitudinal velocity in the beam of
cold atoms which increases with increasing MOT-length. Hence the
total flux should grow with the length of the 2D-MOT. In the limit
of an infinitely long 2D-MOT the longitudinal velocity
distribution of the atom beam becomes equal to a thermal
distribution.\\
Without collisions the number of trappable atoms and thereby the
flux increases with a higher pressure in
the vapor cell. \\
Let us now discuss the effect of collisions. At higher pressures
their evolution creates a thermalization of the atoms in the
volume of the vapor cell. Thereby atoms can now start not only
from the walls but also from within the vapor cell. This increases
the background of thermal atoms in the atomic beam. Moreover
background gas collisions in the vapor cell and light assisted
collisions of excited atoms in the beam with the background gas
are mechanisms that limit the flux when the MOT-length increases.
The mean number of collisions is given by the product of an
ensemble averaged collision rate ${\Gamma = n\,\sigma\,\langle
v\rangle}$ and the mean time ${\tau=\langle z\rangle/v_z}$ which
the atom spends in the cooling volume. $n$ is the density in the
vapor cell, $\sigma$ the collision cross section and ${\langle
v\rangle}$ the mean thermal velocity in the vapor cell.  The
longer the MOT is, the longer becomes the time of propagation
which the atoms spend in the vapor cell. Therefore the higher is
the probability for losses due to collisions such that a further
increase of MOT-length does not produce a higher flux. In this
simple picture the flux is supposed to saturate as a function of
MOT-length. The mean velocity should increase since atoms with
small ${v_z}$ are more vulnerable by collisions. \\
For a given MOT geometry an increasing pressure will decrease the
effective length of the cooling volume. Therefore one should
expect an optimum pressure for a given geometry of the 2D-MOT,
namely when the mean free path in the vapor cell becomes
comparable with the length of the cooling volume. In addition
higher densities in the vapor cell lead to a higher absorption of
the light beams and decrease the
cooling efficiency.\\

For a theoretical description of the flux of cold atoms from a
2D-MOT source we resort to a simple rate model which was
introduced in \cite{wieman90} for vapor-cell MOTs and later
expanded for atomic beam sources in \cite{dieck98}. Based on that
model for the total flux we derive a model for the
longitudinal velocity distribution of the atom flux.\\
We define a function ${\hat{\Phi}}$ which describes the integrated
flux per velocity interval ${[v_z, v_z + dv_z]}$:
\begin{equation}
\hat{\Phi}(n,v_z)=
\frac{\int_0^L\,R(n,v_z,z)\cdot\exp\left(-\Gamma_{coll}(n)\frac{z}{v_z}\right)dz}{1
+ \frac{\Gamma_{trap}(n)}{\Gamma_{out}}}
\end{equation}
Here ${\Gamma_{trap}}$ is the loss rate out of the trapped cloud
due to background gas collisions, ${\Gamma_{out}}$ determines the
outcoupling rate from the captured vapor in the 2D-MOT-trapping
region into the atomic beam. $L$ is the total length of the
cooling volume and $n$ is the density of Rb-atoms in the vapor
cell. $R$ is the loading rate of ${^{87}}$Rb atoms into the
2D-MOT. The effect of light assisted collisions between the
background gas and the cold atoms in the atomic beam on the way to
the tube (described by the collision rate ${\Gamma_{coll}}$) is
implemented by an exponential loss term. In our setup the cooling
region extends directly to the differential pumping tube.
${z/v_z}$ is the time of flight
for the atoms through the MOT-volume.\\
The total flux, which gives the number of atoms per time interval
integrated over the output area of the source, is given by the
integral of ${\hat{\Phi}}$ over all longitudinal velocities:
${\Phi\,=\,\int_0^\infty\,\hat{\Phi}(n,v_z)\,dv_z}$. Only positive
values for ${v_z}$ are taken into account.\\
In order to derive the loading rate for a 2D-MOT, one needs to
consider the flux of atoms through the surface of the cooling
volume. Since we are interested in the loading flux through the
side walls, we need to consider only radial velocities, weighted
with that part of the Boltzmann distribution which is trappable
according to the discussion above. The loading rate is
proportional to the density in the vapor cell and to the surface
area of the cooling volume. For simplicity we take a cylindrical
cooling volume. We define a loading rate per velocity interval
${[v_z, v_z + dv_z]}$:
\begin{eqnarray}
R(n,v_z,z)\,=&\,n\,d\,\frac{16\sqrt{\pi}}{u^3}
\,\exp\left(\frac{-v_z^2}{u^2}\right)\cdot\\
&\int_{0}^{v_c(v_z,z)}v_r^2\,\exp\left(\frac{-v_r^2}{u^2}\right)\,dv_r\nonumber
\end{eqnarray}
Where ${d}$ is the diameter of the cooling volume, ${u =
\sqrt{\frac{2k_BT}{m}}}$ is the most probable velocity of the
Maxwell-Boltzmann-distribution with $k_B$ denoting the Boltzmann
constant and $T$ the temperature of the vapor, ${v_r =
\sqrt{v_x^2+v_y^2}}$ is the radial velocity. The integral's upper
limit ${v_c}$ is the capture velocity which is generally a
function of the longitudinal
velocity ${v_z}$ and of the atom's distance $z$ from the aperture. \\
To satisfy the two asymptotic behaviors at low ${v_z}$ (${v_c
\rightarrow v_{c0}=const}$) and at high ${v_z}$ (${v_c \propto
1/v_z }$) we model ${v_c}$ as:
\begin{equation}
\label{einfanggeschw} v_c(v_z,z) = \frac{v_{c0}}{1 +
\frac{v_z}{v_{cr}}}
\end{equation}
${v_{c0}}$ is the radial capture velocity which lies usually in
the range of 30\,m/s. ${v_{cr}}$ is the so called critical
velocity above which the cooling time is limited by the
longitudinal motion and the capture velocity falls off as
${1/v_z}$. The capture velocity is nearly equal to ${v_{c0}}$
below ${v_{cr}}$. We choose ${v_{cr}}$ via the equality of the
mean longitudinal flight time ${L/(2\,v_{cr})}$ and the radial
cooling time which we approximate by ${d/v_{c0}}$. In this
approximation the explicit $z$-dependence drops out of ${v_c}$.
This results in:
${v_{cr}=L\,v_{c0}/(2\,d)}$.\\
This gives for the flux per velocity interval:
\begin{eqnarray}
\label{flussverteilung} \hat{\Phi}(n,v_z) = &\frac{n d}{1 +
\frac{\Gamma_{trap}(n)}{\Gamma_{out}}}\,\frac{16\,\sqrt{\pi}}{u^3}\,\frac{v_z}{\Gamma_{coll}}\,
\exp\left(\frac{-v_z^2}{u^2}\right)\\
&\left(1-\exp\left(-\Gamma_{coll}\frac{L}{v_z}\right)\right)\,
\int_0^{v_c}v_r^2\exp\left(\frac{-v_r^2}{u^2}\right)\nonumber
\end{eqnarray}
At residual vapor pressures of a few ${10^{-7}}$\,mbar the typical
lifetime of a MOT loaded from atomic beams is about 100\,ms. From
that we conclude a collision rate ${\Gamma_{trap}}$ on the order
of 10\,s${^{-1}}$ for this vapor pressure. As will be discussed in
part \ref{results} the typical cooling time is in the range of ms
therefore the order of magnitude for ${\Gamma_{out}}$ is
${10^{3}}$\,s${^{-1}}$. The collision rate for light assisted
collisions is given by: ${\Gamma_{coll}=n\,\langle
v\rangle\sigma}$. ${\langle v\rangle}$ is the mean velocity in the
vapor, ${\sigma}$ is the effective collision cross section for
light assisted collisions between background gas atoms and atoms
in the cold beam. Following \cite{dieck98,steane} we assume that
this process can be described by resonant dipole-dipole
interaction and follows a ${C_3}$/${R^3}$ potential. $R$ is the
inter-atomic distance.\\
When fitting the experimental results with this theoretical model
we obtain ${\Gamma_{trap}/\Gamma_{out}=\,0.012}$ at a pressure of
${10^{-7}}$\,mbar. This agrees well with the observed lifetimes.
For the effective collision cross section we get: ${\sigma_{eff}
\approx 1.8\cdot 10^{-12}}$\,cm${^2}$. This value matches within a
factor of two with the measurement of Dieckmann et al. \cite{dieck98}. \\
When comparing the theoretical results with the experiment,
equation (\ref{flussverteilung}) needs to be multiplied by an
overall efficiency factor. The fit to our measured data yields an
efficiency factor of ${5\cdot10^{-3}}$. This takes into account
the different intensities, the Gaussian beam profiles and the
absorption in the Rb-vapor. Equation (\ref{flussverteilung})
describes the longitudinal velocity distribution in the cold
atomic beam. A comparison of this model with the measured results
will be given
in part \ref{results}.\\
Let us summarize the expected general behavior:\\
1) An increasing MOT-length should lead to: \\
a) \emph{a higher flux}. Faster atoms can be captured with an
increasing $L$. Due to collisions the expression of
${\hat{\Phi}(n,v_z)}$ becomes independent of $L$ for large values
of $L$. The flux shows a
saturation for lengths above ${L>\frac{\left<v_z\right>}{\Gamma_{coll}(n)}}$.\\
b) \emph{an increasing mean velocity in the atomic beam}. Above an
optimum MOT-length every increase in length will only add faster
atoms to the beam, thus increasing the value for
the mean velocity.\\
2) An increasing density of Rb-atoms in the vapor cell should lead
to: \\
a) \emph{a linear increase of the total flux at low pressures}.
The loading rate is proportional to
${\frac{n}{1+\Gamma_{trap}(n)/\Gamma_{out}}}$ which is linear in
the density for low pressures. At higher pressures the term
${1/\Gamma_{coll}(n)}$, which is inversely proportional to $n$,
dominates. For a given length of the MOT-beams there exists
an optimum pressure above which the flux decreases with increasing pressure.\\
b) \emph{an increasing mean velocity}. The necessary momentum
transfer to be pushed out of the beam is smaller for the slow
atoms, leaving a higher fraction of hotter
atoms in the beam.\\

\section{Experimental Setup and Diagnostics}
\begin{figure}
\begin{center}
\epsfig{file=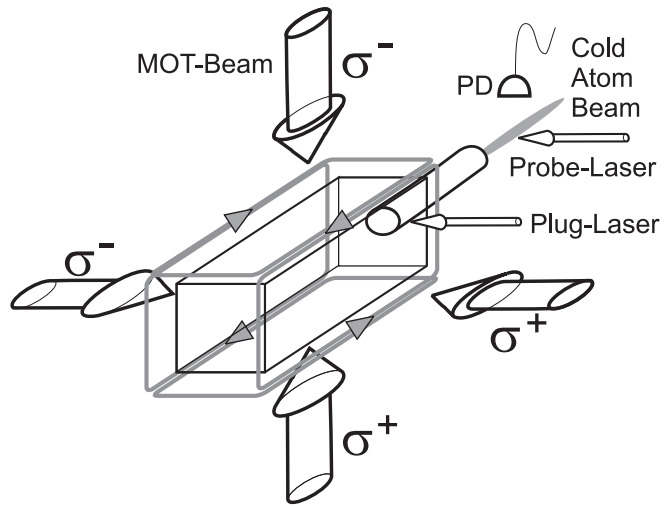,width=\columnwidth} \caption{Schematic view
of the setup. Rectangular coils produce a 2-dimensional magnetic
quadrupole field. The line of zero magnetic field coincides with
the long axis of the glass cell. Four perpendicular laser beams
with circular polarizations in the usual MOT configuration cool
the atoms in two dimensions. The atomic beam travels horizontally
through a differential pumping tube. Analysis of the beam is done
by a time-of-flight-method. A plug beam shuts off the atomic beam
and a probe-laser is shone in perpendicularly to the beam. The
fluorescence is detected by a calibrated photodetector.}
\label{setup}
\end{center}
\end{figure}
\label{experiment}As discussed in the previous section the main
parameters leading to a high flux are the length of the cooling
volume and a high vapor pressure. In our apparatus we
tried to optimize these parameters. \\
Our setup consists of a 2-chamber vacuum-system separated by a
differential pumping tube. The tube is conically shaped, 133\,mm
long and has a diameter of 6\,mm which widens at the UHV-end up to
9.6\,mm. It maintains a pressure drop of three orders of magnitude
between a vapor pressure cell and two UHV-six-way-crosses used as
analyzing chambers. Compared to other setups we use a rather large
aperture for the differential pumping tube. The vapor cell is a
glass cuvette (135mm x 35mm x 35mm) whose long axis (z-axis) is
horizontally aligned. The geometry is chosen in such a way that
the opening angle of the tube does not permit atoms starting on
the side walls to be transmitted without transverse cooling. The
purpose of this tube is mainly to separate
the cold atomic beam from the thermal atoms. \\
Electric heating rods around the glass cell provide homogeneous
and stable heating thus allowing to work at relatively high vapor
pressures between ${10^{-7}}$ and ${3\cdot10^{-6}}$\,mbar. Four
rectangularly shaped, elongated magnetic coils are placed around
the vapor-cell producing a two-dimensional quadrupole field. The
zero magnetic field line is along the axis of the glass cell. We
work with a field gradient of 17\,G/cm. A
schematic view of the setup is depicted in Fig.\,\ref{setup}.\\
Cooling laser light is provided by a Ti:Sapphire-laser. The laser
is red-detuned by 1.9\,$\Gamma$ from the 5S${_{1/2}}$,F=2
$\longrightarrow$ 5P${_{3/2}}$,F=3 transition. To repump atoms
back into the cycling transition an external cavity diode laser is
employed which is stabilized to the 5S${_{1/2}}$,F=1
$\longrightarrow$ 5P${_{3/2}}$,F=2 transition. For the analysis of
the atomic beam another diode-laser is used. This probe-laser is
locked on resonance to the 5S${_{1/2}}$,F=2 $\longrightarrow$
5P${_{3/2}}$,F=3 transition.\\
The light of the cooling laser is split into four separate beams
which are expanded in spherical and cylindrical telescopes placed
in sequence up to a beamsize of roughly 95x15\,mm (horizontal
waist radius w${_z \approx}$ 25\,mm, vertical waist radius
w${_\rho \approx}$ 6\,mm). Two pairs of horizontally and
vertically counterpropagating beams are overlapped in the center
of the glass cell. The repumping light is overlapped with two
horizontal beams. In order to work at high vapor pressures it is
necessary to use four different laser beams. Retroreflection of
the light beams would lead to a strong imbalance in the light
pressure due to the high absorption
in the vapor.\\
The cooling volume extends to the front of the differential
pumping tube. The center of the 2D-MOT is about 40\,mm in front of
the entrance edge of the tube. There is no dark distance which the
atoms travel at background pressure without being transversely
cooled. This upholds the good collimation of the beam until it
leaves the vapor cell.\\

\textbf{Diagnostics}\\
The measurement of the Rb-pressure in the cell is accomplished by
absorption measurement. The frequency of a small laser beam whose
intensity is below saturation is swept across resonance. The
measurement is calibrated by the absorption of Rb-vapor at room
temperature whose vapor pressure is ${10^{-7}}$\,mbar \cite{varian}.\\
Information about the transverse capture velocity of the 2D-MOT
can be obtained by Doppler-spectroscopy perpendicular to the
atomic beam. For that purpose a probe-laser beam with a diameter
of 1\,mm is aligned through the atomic beam orthogonally to its
axis onto a photodiode. When sweeping its frequency across
resonance the Doppler-profile reveals a Gaussian-shape due to the
thermal velocity distribution of the atoms with two dips
symmetrically centered around the maximum. Atoms of the velocity
classes corresponding to the dips have been cooled and therefore
the height of the central part corresponding to the cold atoms
increases. Division by the fitted Gaussian-profile shows this
structure clearly (Fig.\,\ref{radialcapture}). Half of the width
between
the minima corresponds to the capture velocity.\\
The transverse beam profile is investigated in the UHV-part of the
setup. A probe-beam with a diameter of 1\,mm is directed
orthogonally on the atomic beam. Perpendicular to it a CCD-camera
images the fluorescence signal. From the increasing width (FWHM)
of the signal when moving the probe beam along the
z-axis the divergence of the atomic beam is quantified.\\

The analysis of the longitudinal velocity distribution of the
atomic beam is done by a time of flight (TOF) method. A fraction
of the MOT-laser intensity is split apart and shone into the vapor
cell directly in front of the tube perpendicular to the atomic
beam axis. This beam has a diameter of $\approx$ 8\,mm, and its
intensity was held at 150\,mW throughout all measurements. It
deflects all atoms with a longitudinal velocity lower than
130\,m/s and hence plugs the atomic beam for all smaller
velocities. After a flight-distance of 145\,mm a light-sheet of
1\,mm width from a probe laser is irradiated on the atomic beam in
the UHV-chamber orthogonal to the atom beam. A fraction of the
repumping light is overlapped with the probe-laser. A calibrated
photodiode detects the fluorescence of the atoms. After plugging
the atomic beam the fluorescence signal fades out. From this
signal information about the velocity distribution of the atomic
flux can be derived according to:
\begin{equation}
\Phi(v_z) = \frac{\eta}{d_{probe}} \frac{l}{v_z} \frac{dS}{dt}
\end{equation}
Where $S$ is the signal from the photodiode, ${d_{probe}}$ the
width of the probe-beam light-sheet, $l$ the distance between
plug-beam and probe-beam and $\eta$ contains calibration
parameters of the detection system.\\

\section{Experimental Results}
\label{results} 
\begin{figure}[h]
\begin{center}
\epsfig{file=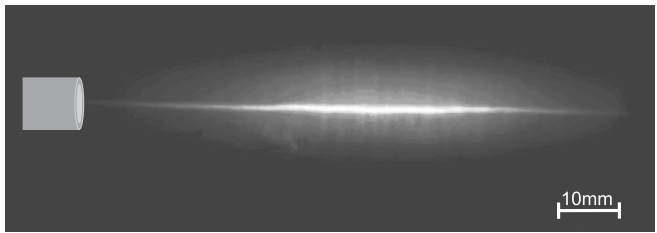,width=\columnwidth} \caption{Picture of the
fluorescence in the 2D-MOT with a flux of
${3.5\cdot10^{10}}$\,atoms/s. The vapor pressure in the glass cell
is about ${7\cdot10^{-7}}$\,mbar. A beam of approximately 90\,mm
length and a thickness of 2\,mm develops. The cooling region
extends to the differential pumping tube which is sketched on the
left.} \label{2dmotpic}
\end{center}
\end{figure}

\begin{figure}[h]
\begin{center}
\epsfig{file=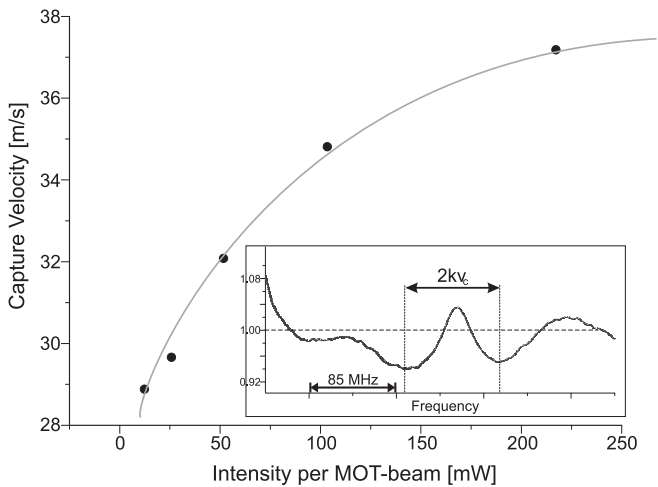,width=\columnwidth} \caption{Transverse
capture velocity of the 2D-MOT. The inset shows the normalized
Doppler-spectroscopy signal, i.e. the measured signal divided by a
fitted Gaussian distribution. The width between the two minima
corresponds to twice the capture velocity. The main graph shows
the dependence of the capture velocity on the intensity in the
cooling laser beams. The line serves merely to guide the eye. The
capture velocity saturates at high intensities at a value of
38\,m/s.} \label{radialcapture}
\end{center}
\end{figure}

At first we give information about the
characteristics of the 2D-MOT: its loading time and its capture
velocity. This is followed by a discussion of the properties of
the atomic beam. The influence of laser power, length of the
cooling volume and of the pressure in the vapor cell is studied.\\
The two-dimensional laser-cooling produces a about 2\,mm thin,
90\,mm long, line of high intensity of fluorescence light in the
glass cell along the axis of zero magnetic field which is clearly
visible at low vapor pressures. A picture of the fluorescence
light in the
2D-MOT is shown in Fig.\,\ref{2dmotpic}.\\
A measurement of the radial capture velocity in the above
described way yields the maximum transverse capture velocity
${v_{c0}}$ in equation (\ref{einfanggeschw}). The capture velocity
depends merely on the intensity of the laser beams, the detuning
and on the magnetic field gradient. Figure\,\ref{radialcapture}
displays the dependence of ${v_{c0}}$ on the laser intensity. The
inset shows the Doppler-spectroscopy signal from which the capture
velocity is inferred. In the intensity range which we apply, it
extends from 28\,m/s to 38\,m/s. For high laser powers above
160\,mW per beam (which corresponds to an intensity of
$\simeq$17\,mw/cm${^2}$) it saturates at a value of 38\,m/s. This
value matches well with an estimation when equating the frequency
shift due to the Zeeman effect with the detuning of the laser,
which gives the linear
capture range - in our case 35\,m/s.\\

For measuring the loading time of the 2D-MOT we directed a 1\,mm
thick probe laser beam through the center of the atomic beam onto
a photodiode. When switching on the 2D-MOT cooling light the
absorption decreased abruptly because the F=2
${\longrightarrow}$F=3-transition is driven by the MOT laser. The
absorption increases again and reaches a steady state when the
atoms are transversely cooled and closer in resonance with the
probe laser than with the MOT light. The ${1/e}$-time of this
increase is about 2\,ms. This gives the characteristic time scale
for the cooling process. From that we can estimate the value for
the outcoupling rate ${\Gamma_{out}}$ in equation
(\ref{flussverteilung}) to be on the order of ${10^3}$\,s${^{-1}}$.\\
The emerging atomic beam is well collimated. From our transverse
beam profile measurement we deduce a beam-divergence of 32\,mrad.
This is about a factor of 2 less than the geometrically allowed
divergence by the differential pumping tube of 59\,mrad. This
means that the beam of cold atoms is not hindered by the aperture.
It is possible to further suppress the thermal background by
diminishing the aperture of the differential pumping tube without
decreasing the flux of cold atoms in future experiments.\\

\begin{figure}[h]
\begin{center}
\epsfig{file=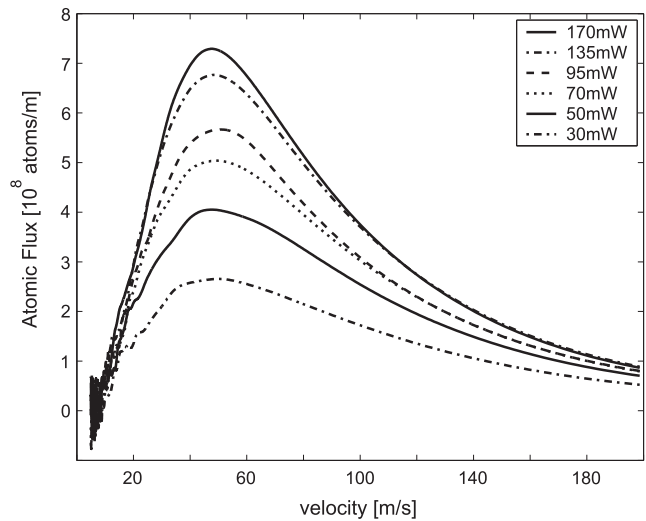,width=\columnwidth} \caption{Distribution of
the atomic flux versus the longitudinal velocity. The laser power
in the cooling beams was varied from 30\,mW to 170\,mW. The
pressure for this measurement is ${1.6\,10^{-6}}$\,mbar, the
length is 92\,mm. A small shift of the mean velocity to higher
values with increasing laser power is visible. The velocity
distribution is centered around 50\,m/s and has a width of about
75\,m/s.} \label{fluxoverpower}
\end{center}
\end{figure}
\begin{figure}[h]
\begin{center}
\epsfig{file=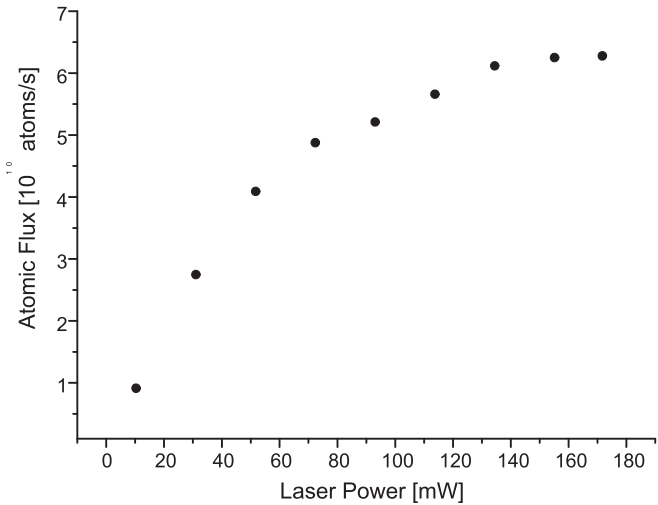,width=\columnwidth} \caption{Dependence of
the total atomic flux on the cooling laser power per beam. The
total flux in atoms/s is given by the area under the curves in
graph \ref{fluxoverpower}. It ranges from ${1\cdot10^{10}}$ to
${6\cdot10^{10}}$ atoms/s. The total flux depends strongly on
laser power and saturates at values of 160\,mW per laser beam.}
\label{powerdep}
\end{center}
\end{figure}

\begin{figure}[h]
\begin{center}
\epsfig{file=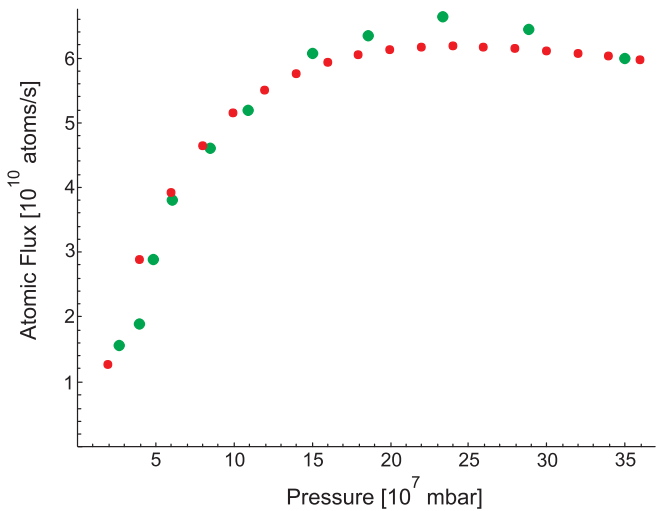,width=\columnwidth} \caption{Dependence of
the atomic flux on the Rb-pressure in the vapor cell. The dark
spots mark the theoretical results whereas the bright circles
describe the experimental results. The measurement was done at
full length of the MOT laser beams (92\,mm) and at a laser power
of 170\,mW. At low pressures the atomic flux increases linearly
whereas it finds a maximum at about 2$\cdot$10$^{-6}$\,mbar and
decreases for higher values. The mean free path in the cell
reaches the value of the length of the MOT at the pressure for the
maximum flux - a clear hint that collisions limit a further
increase of the atomic flux. The theoretical model agrees well
with the experimental data.} \label{fluxoverpressure}
\end{center}
\end{figure}

The time-of-flight measurements give information about the
distribution of the atomic flux versus velocity. A typical set of
velocity distributions is depicted in Fig.\,\ref{fluxoverpower}.
The varied parameter for the single curves is the power in the
laser beams. We observe a relatively broad feature with a peak
velocity of around 50\,m/s and a width of roughly 75\,m/s. An
increase of the peak velocity with increasing laser power is
visible. The transverse cooling works more efficiently with
increasing laser intensity. Therefore atoms with a higher velocity
in the z-direction contribute to the beam. The integral of the
flux-distribution gives the total flux of atoms per unit time.
This is displayed in Fig.\,\ref{powerdep}. The total flux
saturates for laser powers above 160\,mW per laser beam. We
observe a maximum flux of 6$\cdot$10${^{10}}$\,atoms/s at a laser
power of 160\,mW per laser
beam and at a vapor pressure of ${1.8\cdot10^{-6}}$\,mbar. \\
\begin{figure}[h]
\begin{center}
\epsfig{file=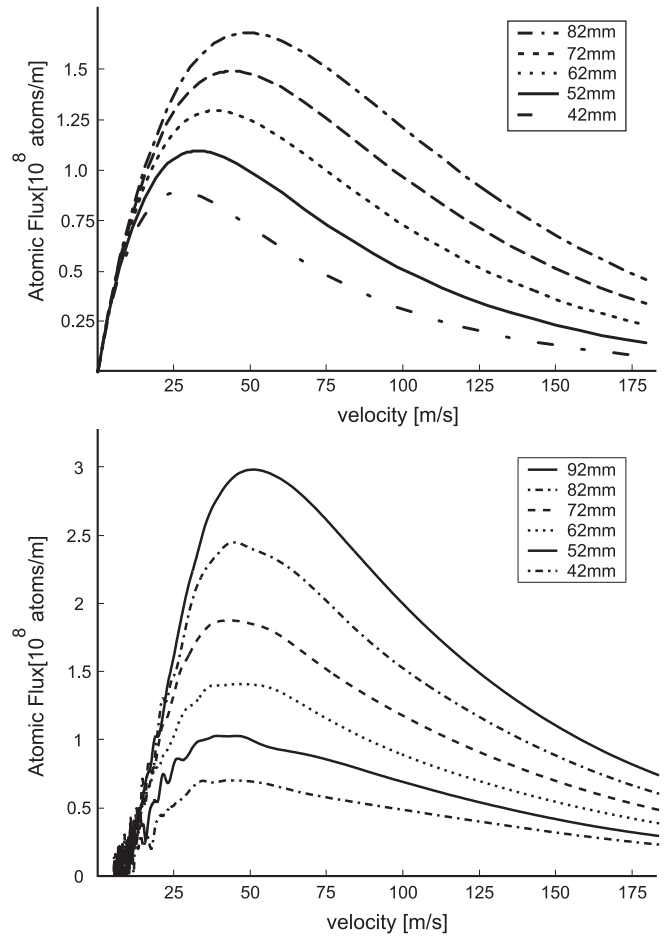,width=\columnwidth} \caption{Dependence of
the atomic flux on the length of the MOT-beams. The upper graph
shows the prediction by our model. The experimental result is
plotted in the lower graph. The power per laser beam is held
constant at 21\,mW and the Rb-pressure is
1.6$\cdot$10$^{-6}$\,mbar. The peak of the velocity distribution
is shifted to higher values for a longer cooling volume. The
atomic flux increases with the length of the cooling volume. In
this measurement an increasing part of the MOT-beams is blocked
while the laser power is adjusted so that the total laser power
incident on the atoms stays constant.} \label{lengthdep}
\end{center}
\end{figure}
In addition to the field gradient and the detuning, the size of
the MOT-beams determines the capture range of the MOT. As
discussed in part \ref{theory} the length of the cylindrical
MOT-beams strongly influences the flux of cold atoms.
Fig.\,\ref{lengthdep} shows the distribution of the atomic flux
versus velocity when the length of the beams is varied. The upper
part shows the result of our theoretical model. The lower part
displays the TOF-results. This measurement was done by
successively blocking a part of all four cylindrical MOT-beams
starting on the back side of the glass cell while simultaneously
increasing the laser power per beam. Thus the total power shining
on the atoms is kept constant and it is ensured that we see a pure
influence of the MOT-length. With increasing length the flux grows
and the maximum velocity shifts to higher values.
\begin{figure}[h]
\begin{center}
\epsfig{file=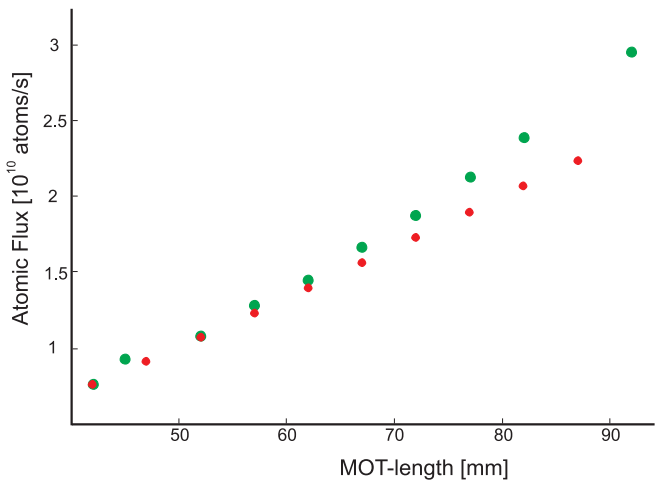,width=\columnwidth} \caption{Dependence of
the total flux on the length of the cooling volume. The
experimental parameters are the same as in Fig.\,\ref{lengthdep}.
The dark spots show the theoretical the bright circles the
experimental data.} \label{fluxoverlength}
\end{center}
\end{figure}
Fig.\,\ref{fluxoverlength} shows the total flux as a function of
the MOT length. The total flux is expected to saturate for a
cooling volume above an optimum length which is given by the mean
velocity and the collision rate ${\Gamma_{coll}(n)}$ as discussed
in part \ref{theory}. However, for the pressure range of this
measurement no saturation of the flux is visible even at the
maximum MOT-length of 92\,mm. The discussion of the geometry
dependence of the total flux clearly reveals the advantage of
utilizing the whole laser power across a long cooling volume
compared to
increasing the laser intensity in a usually sized MOT.\\
The longer the MOT the longer is the interaction region for the
atoms to be transversely cooled. Therefore also higher velocity
classes can be gathered into the beam. This explains that the
maximum velocity is shifted towards higher values with
increasing MOT-length. \\
With an increasing length the mean velocity becomes larger. For an
infinitely long 2D-MOT and low vapor pressures the velocity
distribution should approach a thermal distribution. In
applications like loading a 3D-MOT, the choice of the 2D-MOT
length is limited by the 3D-MOT's maximum capture
velocity.\\
Our model is in good agreement with the measurements and predicts
the essential features that we observe. Only the width of the
velocity distribution is not accurately predicted. \\
\begin{figure}[h]
\begin{center}
\epsfig{file=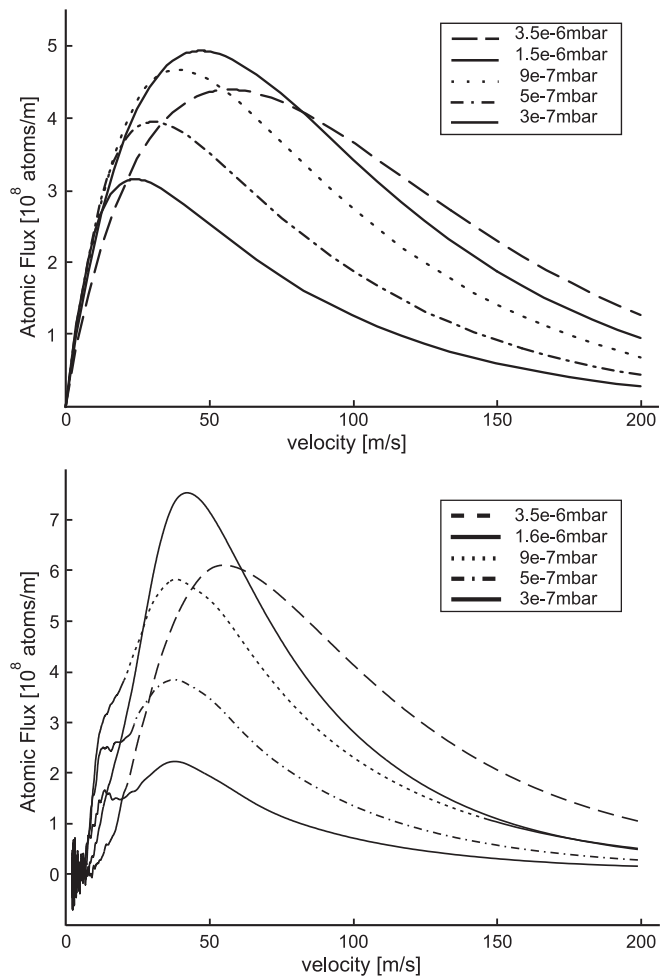,width=\columnwidth} \caption{Distribution of
the atomic flux versus the longitudinal velocity. The pressure in
the vapor cell was varied from ${10^{-7}}$\,mbar to
${3.5\cdot10^{-6}}$\,mbar. The upper graph shows the results of
our model. The experimental results are plotted in the lower
graph. The model describes an increase in the peak velocity with
increasing pressure and also a growing flux with higher pressures
which is in accordance with the experimental results. Above an
optimum pressure the flux decreases again. This is also visible in
the dashed line both in the model and in the measurement.}
\label{pressuredep}
\end{center}
\end{figure}
The variation of the Rb-vapor pressure is done by changing the
temperature of the glass cell in a controlled way. It is possible
to raise the vapor pressure from 1$\cdot$10$^{-7}$\,mbar to
3$\cdot$10$^{-6}$\,mbar. Fig.\,\ref{pressuredep} demonstrates the
dependence of the velocity distribution on the vapor pressure. The
upper part shows the behavior as described by our theoretical
model, the lower graph displays the TOF-measurements. The flux
increases with increasing pressure, reaches a maximum at
1.5$\cdot$10$^{-6}$\,mbar and decreases again for higher
pressures. In the high pressure regime the mean free path of
Rb-atoms in the vapor cell is on the order of a few cm and is
comparable to the dimensions of the atomic beam in the 2D-MOT.
This means that collisions start to limit the atomic flux as is
discussed in part \ref{theory}. Above the optimum pressure value
the total flux decreases again. The optimum pressure depends on
the length of the 2D-MOT. The longer the 2D-MOT the smaller is the
value for the optimum pressure. Our value for the optimum pressure
agrees well with the prediction of Vredenbregt et al \cite{vredenbregt}.\\
The maximum velocity of the distribution shifts towards higher
values with increasing pressures. This is revealed in the
experiment and also in the theoretical curves. Atoms with a small
longitudinal velocity are more vulnerable by collisions because a
small transverse momentum transfer already produces a large enough
divergence so that the atoms collide with the tube. A higher
pressure leads to a larger mean velocity in the atomic beam.\\
The total flux as a function of vapor pressure is shown in
Fig.\,\ref{fluxoverpressure}. The linear increase of the flux at
low pressures and the existence of an optimum pressure is well
described by our model.\\
\begin{figure}[h]
\begin{center}
\epsfig{file=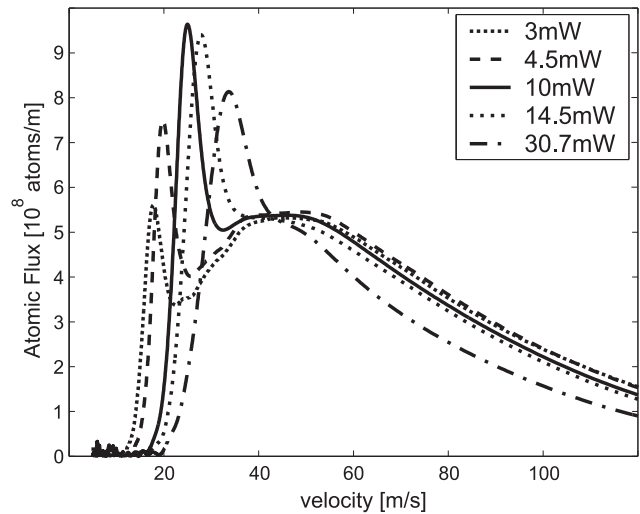,width=\columnwidth } \caption{When a laser
beam is shone on the axis copropagating with the atomic beam, the
longitudinal velocity distribution changes dramatically. A narrow
peak rises up at velocities between 20\,m/s and 40\,m/s. Its width
and position on the ${v_z}$-axis depends on the intensity in the
pushing beam. Atoms at very low velocities below 15\,m/s are
pushed out of the beam or are accelerated. The total flux stays
nearly constant.} \label{axialbeams}
\end{center}
\end{figure}
In addition to the Doppler-cooling laser light we shone in a laser
beam of the same detuning on the axis copropagating with the atoms
(pushing beam). This setup comes close to the 2D${^+}$-MOT of
Dieckmann et al \cite{dieck98}. The new longitudinal velocity
distribution is shown in Fig.\,\ref{axialbeams}. A narrow (width
$\approx$7.5\,m/s) and intense peak at low velocities (centered
around 25\,m/s) rises. The width of this feature increases and its
position is shifted towards higher values when the power in the
pushing beam is increased. Since there is no magnetic field
gradient on the axis, the axial beam addresses only a certain
velocity class of atoms. The data indicates that a group of atoms
propagating in the negative z-direction is slowed down and their
direction of propagation is turned around into the positive
z-direction. At the same instance the very slow atoms are pushed
out of the beam. The velocity distribution shows almost no atoms
at velocities below 15\,m/s. The total flux stays nearly constant
when shining in the pushing beam. Only for high powers (above
15\,mW) in the pushing beam the 2D-MOT is too much disturbed and
the flux decreases as is shown by the curve for 30.7\,mW in
Fig.\,\ref{axialbeams}. A different detuning in the axial beams
from the usual MOT-laser beams could address the high velocity
classes and increase the number of slow atoms in the beam. This
slowing effect can be used to increase the flux at low velocities
which might be useful when loading conventional 3D-MOT
configurations with
capture ranges around 30\,m/s.\\
We verified the results of the TOF technique with the method of
Doppler-spectroscopy directly on the atomic beam. For that purpose
another light sheet from the probe laser was aligned
counterpropagating to the atomic beam at an angle of 3.5${^\circ}$
to it. When scanning the laser-frequency one obtains
Doppler-profiles which confirm the longitudinal velocity profiles
of the beam obtained by the TOF-measurements. In this measurement
we could verify that the thermal background transmitted by the
differential pumping tube is negligible compared to the high flux
of cold atoms.\\

\section{Summary and Outlook}
\label{outlook} We have realized and investigated a novel setup
for an intense source of slow atoms. The underlying physics is
pure two-dimensional cooling and trapping. To reach a high flux
with a small divergence we work at high vapor pressures and with a
long cooling region. The length of the 2D-MOT provides good
collimation (divergence $\approx$ 32\,mrad) and a mean velocity of
50\,m/s. The high vapor pressure assures a large loading rate and
thus a high flux. To be able to work at pressures in the range of
${10^{-6}}$\,mbar we work with four separate counterpropagating
cooling laser beams. The total flux is in the range of several
${10^{10}}$ atoms per second. A maximum flux of ${6\cdot10^{10}}$
atoms/s was obtained at a 2D-MOT length of 90\,mm, laser intensity
of 160\,mW per beam
and a vapor pressure in the glass cell of ${1.6\cdot10^{-6}}$\,mbar.\\
This alternative source of atoms provides a total flux comparable
to a Zeeman slower with less total length, less material
throughput and without disturbing light and magnetic fields for a
consecutive MOT in a UHV system and therefore provides a very good
starting point for further laser and evaporative cooling
experiments and BEC-generating apparatus \cite{beamevap,makingprobing}.\\
In a consecutive experiment we loaded the beam into a 3D-MOT with
a capture velocity of about 50\,m/s which traps a large fraction
of the atoms.\\
\section*{Acknowledgement}
This work was supported by the Schwerpunktprogramm:
``Wechselwirkung in ultrakalten Atom- und Molek\"ulgasen'' (SPP
1116) of the Deutsche Forschungsgemeinschaft and by the European
Research Training Network: ``Cold Quantum Gases'' under the
contract No: HPRN-CT-2000-00125. \\


%

\newpage











%


%
%


\begin{thebibliography}{99}
\bibitem{sigel94} C.S. Adams, M. Sigel and J. Mlynek,
Phys. Rep. \textbf{240}, 143 (1994).\\

\bibitem{metcalf} H.J. Metcalf, P. van der Straten, \textit{Laser Cooling
and Trapping}, Springer, New York (1999), ISBN 0-387-98728-2.\\
\bibitem{ketterledruten} W. Ketterle, N.J. van Druten, Evaporative
Cooling of Trapped Atoms, Adv. At. Mol. Opt. Phys., \textbf{37},
181 (1996).\\
\bibitem{zeemanslower} W. Phillips and H.J. Metcalf, PRL.
\textbf{48}, 596-599 (1982).\\
\bibitem{LVIS96}  Z.T. Lu, K.L. Corwin, M.J. Renn, M.H. Anderson,
E.A. Cornell, and C.E. Wieman, Phys. Rev. Lett. \textbf{77}, 3331
(1996).\\
\bibitem{riischu} E. Riis, D.S. Weiss, K.A. Moler, S. Chu, Phys. Rev. Lett.
\textbf{64}, 1658 (1990).\\
\bibitem{swanson} T.B. Swanson, N.J. Silva, S.K. Mayer, J.J. Maki,
D.H. McIntyre, JOSA B \textbf{13}, 9, 1833 (1996).\\
\bibitem{riis} H. Chen, E. Riis, Appl. Phys. B \textbf{70}, 665 (2000).\\
\bibitem{nellessen} J. Nellessen, J. Werner, W. Ertmer, Opt. Comm. \textbf{78},
300 (1990).\\
\bibitem{Yu} J. Yu, J. Djemaa, P. Nosbaum, P. Pillet, Opt. Comm. \textbf{112},
136 (1994).\\
\bibitem{berthoud} P. Berthoud, A. Joyet, G. Dudle, N. Sagna, P. Thomann, Europhys. Lett.
\textbf{41}, 141 (1998).\\
\bibitem{dieck98}  K. Dieckmann, R.J.C. Spreeuw, M.
Weidem{\"u}ller and J.T.M. Walraven, Phys. Rev. A \textbf{58},
3891 (1998).\\
\bibitem{camposeo} A. Camposeo, A. Piombini, F. Cervelli, F. Tantussi, F. Fuso, E. Arimondo,
Opt. Comm. \textbf{200}, 231 (2001).\\
\bibitem{wieman90} C. Monroe, W. Swann, H. Robinson and C. Wieman,
Phys. Rev. Lett. \textbf{65}, 1571 (1990).\\
\bibitem{steane} A.M. Steane, M. Chowdhury and C.J. Foot, JOSA B,
\textbf{9}, 2142 (1992).\\
\bibitem{vredenbregt} E.J.D. Vredenbregt, K.A.H. van Leeuwen,
H.C.W. Beijerinck, Opt. Comm. \textbf{147}, 375 (1998).\\
\bibitem{varian} M.H. Hablanian, \textit{High-Vacuum Technology}, M.
Dekker, New York (1990), ISBN 0-8247-8197-X.\\
\bibitem{makingprobing} W. Ketterle, D.S. Durfee, and D.M.
Stamper-Kurn, Proceedings of the International School of Physics
"Enrico Fermi", Course CXL, IOS Press, Amsterdam, 1999) pp.
67-176.\\
\bibitem{beamevap} E. Mandonnet, A. Minguzzi, R. Dum, I.
Carusotto, Y. Castin and J. Dalibard, Eur. Phys. J. D \textbf{10},
9-18 (2000).\\


\end{thebibliography}
\end{document}